\documentstyle[mnras_cite,onecolumn]{mymn}

\newcommand{\ltappeq}{\raisebox{-0.6ex}{$\,\stackrel
{\raisebox{-.2ex}{$\textstyle <$}}{\sim}\,$}}
\newcommand{\gtappeq}{\raisebox{-0.6ex}{$\,\stackrel
{\raisebox{-.2ex}{$\textstyle >$}}{\sim}\,$}}

\newcommand{\be}{\begin{equation}}
\newcommand{\ee}{\end{equation}}
\newcommand{\ba}{\begin{eqnarray}}
\newcommand{\ea}{\end{eqnarray}}

\begin{document}

\title[Steady-state, mass-losing accretion disks]{The radial effective 
temperature distribution of steady-state, mass-losing accretion disks} 

\author[C. Knigge] {Christian Knigge$^*$\\ 
Department of Astronomy, 
Columbia University, 
550 West 120th Street,
New York, NY 10027, USA}

\date{Accepted 0000-00-00. Received 0000-00-00}

\def\LaTeX{L\kern-.36em\raise.3ex\hbox{a}\kern-.15em
    T\kern-.1667em\lower.7ex\hbox{E}\kern-.125emX}

\let\leqslant=\leq 

\newtheorem{theorem}{Theorem}[section]

\maketitle
\footnotetext{$^*$Hubble Fellow}

\begin{abstract}

Mass loss appears to be a common phenomenon among disk-accreting
astrophysical systems. An outflow emanating from an accretion disk
can act as a sink for mass, angular momentum and energy and can 
therefore alter the dissipation rates and effective temperatures
across the disk. Here, the radial distributions of dissipation rate
and effective temperature across a Keplerian, steady-state,
mass-losing accretion disk are derived, using a simple,
parametric approach that is sufficiently general to be applicable to 
many types of dynamical disk wind models.

Effective temperature distributions for mass-losing accretion disks in
cataclysmic variables are shown explicitly, with parameters chosen 
to describe both radiation-driven and centrifugally-driven outflows. 
For realistic wind mass-loss rates of a few percent, only
centrifugally-driven outflows -- particularly those in which mass
loss is concentrated in the inner disk -- are likely to alter the disk's
effective temperature distribution significantly. Accretion disks that 
drive such outflows could produce spectra and eclipse light curves that
are noticeably different from those produced by standard, conservative 
disks. 

\end{abstract}

\section{Introduction}
\label{demo:intro}

\indent Accretion disks are an important ingredient in our current 
understanding of many astrophysical systems on all
scales. Examples of presumed disk accretors include young
stars, compact objects in close binary systems and active 
galactic nuclei (AGN) and quasars (QSOs). There is 
also evidence that the process of mass accretion via a 
disk is often -- and perhaps always -- associated with mass-loss 
from the disk in the form of a wind or a jet. For instance, 
the broad-line region in QSOs, the radio jets in AGN, the jets 
and bipolar outflows from young stars 
and the fast winds observed in disk-accreting cataclysmic variables
all provide compelling evidence for such a disk-wind connection
[see Livio (1999) for a recent review].
\nocite{livio7} 

\indent Since the outflow provides an additional sink for mass,
angular momentum and energy, the spectrum emitted by a mass-losing
accretion disk will in general be different from that produced by a
``conservative'' one. Intuitively, mass loss may be expected to reduce
the effective 
temperatures across the entire disk, but especially in those regions
from which matter is preferentially expelled. Based on this expectation,
it has been suggested, for example, that mass loss from the inner disk
may be responsible for the apparently flatter-than-expected radial
brightness temperature distributions inferred for some disk-accreting
nova-like CVs \cite{rutten1}.

\indent Despite this, few attempts have been made to generalize
steady-state accretion disk theory to include the family of
mass-losing disks. As a result, the standard $T \propto R^{-3/4}$ law
that is valid for conservative disks is almost invariably used in practice
(i.e. in modeling and in comparisons between theory and observations),
even in the context of disk-accreting systems in which mass loss could 
alter the disk's effective temperature distribution significantly. Where 
modifications to the standard picture have been attempted, these have
usually been {\em ad hoc} (e.g. Shlosman, Vitello \& Mauche 1996).
\nocite{shlos2} An exception is the work by \scite{piran1}, which provides 
analytic expressions (in integral form) for the energy generation rate 
in a mass-losing accretion disk. However, the bulk of 
Piran's paper deals specifically with thermally-driven, evaporative 
disk winds. The only other type of mass-loss considered explicitly by 
him is a ``delta-function outflow'' from a particular radius on the 
disk. 
\nocite{cann1,bland1,pell1,proga1,emmer1,feld1,feld2,murray2}

\indent The purpose of the present work is twofold. First, simple, 
analytical expressions are derived for the radial distributions 
of dissipation rate (Section~\ref{secdiss}) and effective temperature 
(Section~\ref{tofr}) across a steady-state accretion disk that is 
losing mass and angular momentum to an outflow from its surface. 
The derivation is designed to closely parallel standard, steady-state 
accretion disk theory, as described, for example, by Pringle (1981).
The parametric model adopted to describe the mass loss is simple, 
yet sufficiently general to be applicable to many types of dynamical 
disk wind models, including both radiation-driven outflows 
(e.g. Murray \& Chiang 1996; Proga, Stone \& Drew 1998; 
Feldmaier \& Shlosman 1999; Feldmaier, Shlosman \& Vitello 1999) and 
centrifugally-driven, magneto-hydrodynamic (MHD) disk winds 
(e.g. Blandford \& Payne 1982; Cannizzo \& Pudritz 1988; 
Emmering, Blandford \& Shlosman 1992; Pelletier \& Pudritz 1992). 

Second, the formalism is applied to accretion disks in cataclysmic 
variable stars (CVs; Section~\ref{cvs}). The mechanism responsible for 
the mass loss from these systems is currently still unknown, so both 
radiation-driven (Section~\ref{rad-drive}) and centrifugally-driven 
outflows (Section~\ref{mag-drive}) are considered as possible culprits. 
It is found that if CV winds are of the latter type, they could leave 
an observable imprint on the disk's temperature profile.

\section{The viscous dissipation rate in a steady-state, mass-losing disk}
\label{secdiss}

\indent It is convenient to start by defining the cumulative mass-loss
rate from the disk as  
\be 
\dot{M}_w(R) = \int^R_{R_*} \dot{m}_w(R^\prime) \; 4\pi R^\prime \; dR^\prime.
\label{loser}
\ee
Here $R_*$ denotes the radius at the inner edge of the disk
($R_* \simeq R_{WD}$ in non-magnetic CVs, where $R_{WD}$ corresponds
to the radius of the white dwarf [WD] at the centre of the disk) and
$\dot{m}_w(R)$ is the mass-loss rate per unit area from each disk
face. With this definition the total mass-loss rate is given by 
\be
\dot{M}_{w,total}=\dot{M}_w(R_{disk}),
\ee
where $R_{disk}$ is the radius of the accretion disk. The radial rate
of change of $\dot{M}_w(R)$ is just 
\be 
\frac{\partial \dot{M}_w(R)}{\partial R} = 4 \pi R \; \dot{m}_w(R).
\ee

\indent The equations of mass and angular momentum conservation for a
thin disk subject to mass and angular momentum loss can now be derived 
in close analogy to the usual case by considering a particular annulus
on the disk with inner radius $R$ and outer radius $R+\Delta R$
\cite{pringle1,frank1}. 

The mass of this annulus is $2 \pi R
\Delta R \Sigma$ -- where $\Sigma$ is the surface density of the disk --
and the time rate of change of this is equal to the rate at which mass
flows into the annulus from its radial borders minus the rate at which
it loses mass to the wind. In the limit $\Delta R \rightarrow 0$, mass
conservation then gives
\be
\frac{\partial \Sigma}{\partial t} =
\frac{-1}{R}\frac{\partial}{\partial R}(R \Sigma V_R) - \frac{1}{2 \pi
R} \frac{\partial \dot{M}_w}{\partial R},
\label{masscons}
\ee
where $V_R$ is the radial inflow velocity of material in the disk
($V_R < 0$).

\indent In order to write down the angular momentum equation, it is
necessary to specify how, and how much, angular momentum is lost to 
the wind. Here, it will be assumed that matter ejected at radius $R$ on
the disk carries away specific angular momentum $(lR)^2\Omega$, where
$\Omega$ is the angular velocity in the disk at $R$. 
Thus $l=0$ corresponds to a non-rotating disk wind and $l=1$ to outflowing 
material which retains the specific angular momentum it had at the point
of ejection. Models in which the lever-arm $l > 1$ effectively 
describe an outflow with bead-on-a-wire rotation out to a radius
$R_{out} = l R$, as envisaged for centrifugally-driven magnetic disk
winds. Models such as this can remove a lot of angular momentum from
the disk. 

\indent With the angular momentum carried by the outflowing material
thus fixed, conservation of angular momentum gives 
\be
\frac{\partial \Sigma R^2 \Omega}{\partial t} =
\frac{-1}{R}\frac{\partial}{\partial R}(R^3 \Sigma V_R \Omega) + 
\frac{1}{R}\frac{\partial}{\partial R}(R^3 \nu \Sigma \frac{\partial
\Omega}{\partial R}) - 
\frac{(lR)^2 \Omega}{2 \pi R} \frac{\partial \dot{M}_w}{\partial R}.
\label{angcons}
\ee
The first two terms on the right hand side of this expression
describe the inflow of angular momentum through the boundaries of the
annulus and the effects of viscous torques due to shear ($\nu$ here is
the effective kinematic viscosity). The third term allows for the
angular momentum sink provided by the outflow. Expressions closely 
analogous to Equations~(\ref{masscons}) and (\ref{angcons}) were first 
written down by \scite{bath4}, who examined the process of stream 
penetration, i.e. the effects on the disk of a radially varying mass 
input with constant specific angular momentum.

\indent To keep the discussion simple, the disk will now be assumed to
be Keplerian and in a steady-state. The first of these assumptions gives 
$\Omega (R) = (GM_*/R^3)^{1/2}$, where $M_*$ is the mass of the
accretor, whereas the second implies that $\frac{\partial}{\partial
t}=0$ in~(\ref{masscons}) and~(\ref{angcons}). The conservation
equations can then be integrated, which for~(\ref{masscons}) gives   
\be 
-2 \pi R \Sigma V_R - \dot{M}_w(R) = {\rm constant}.
\label{masscons2}
\ee
The first term on the left hand side of this expression 
is clearly the accretion rate at radius $R$ on the disk,
$\dot{M}_{acc}(R)$. The constant of integration in~(\ref{masscons2})
can be found by considering the boundary condition at 
$R=R_*$, which shows that it is simply the rate of accretion onto the
central star, i.e.  
\be
\dot{M}_{acc}(R) = \dot{M}_{acc}(R_*) + \dot{M}_{w}(R).
\label{masscons3}
\ee

\indent The angular momentum integral is somewhat more difficult to
compute, since as a result of the factor of $R^2 \Omega$ in the new 
outflow sink term, the right hand side of Equation~(\ref{angcons}) is no
longer a perfect differential. That term, at least, must therefore be
integrated explicitly, which requires $\dot{M}_w(R)$ or equivalently
$\dot{m}_w(R)$ to be specified. To keep things tractable, a simple
power law form will be used here, 
\be 
\dot{m}_w(R) = K R^{\xi},
\ee
with the constant $K$ being fixed by normalizing to $\dot{M}_{w,total}$ 
(which is used as a free parameter). Note that, in principle, the local
mass-loss rate may itself depend on the local effective temperature;
in this case an iterative procedure would be needed to deal with the 
resulting non-linearity. 

\indent Equation~(\ref{angcons}) can now be integrated
straightforwardly. Using the usual boundary 
condition, $\frac{\partial \Omega(R_*)}{\partial R}=0$ (which really
applies at or near the boundary layer, rather than at the stellar
surface itself; e.g. Frank, King \& Raine 1985), and the
identification $-2 \pi R \Sigma V_R = \dot{M}_{acc}(R)$ made above for
the mass conservation integral, this yields
\be
3 \pi \nu \Sigma  =  \left\{ \begin{array}{ll} 

\dot{M}_{acc}(R) 
- \dot{M}_{acc}(R_*)\left(\frac{R_*}{R}\right)^{1/2} 
+ \dot{M}_{w,total}
\left(\frac{R_*}{R}\right)^{1/2}
\left[\frac{l^2 (\xi + 2)}{(\xi + 5/2)}\right]
\left[\frac{1-\left(\frac{R}{R_*}\right)^{(\xi+5/2)}}
{\left(\frac{R_{disk}}{R_*}\right)^{(\xi+2)}-1}\right] 
& \xi \neq -2,-5/2 \\ \\

\dot{M}_{acc}(R) 
- \dot{M}_{acc}(R_*)\left(\frac{R_*}{R}\right)^{1/2} 
+ \dot{M}_{w,total} 
\left(\frac{R_*}{R}\right)^{1/2}
[2 l^2]
\left[\frac{1-\left(\frac{R}{R_*}\right)^{1/2}}
{\ln{\left(\frac{R_{disk}}{R_*}\right)}}\right]
& \xi = -2 \\ \\ 

\dot{M}_{acc}(R) 
- \dot{M}_{acc}(R_*)\left(\frac{R_*}{R}\right)^{1/2}
+ \dot{M}_{w,total}
\left(\frac{R_*}{R}\right)^{1/2}
\left[\frac{l^2}{2}\right]
\left[\frac{\ln{\left(\frac{R_*}{R}\right)}}
{1-\left(\frac{R_{*}}{R_{disk}}\right)^{1/2}}\right]
& \xi = -5/2.
\end{array} \right.
\label{bloodylong}
\normalsize
\ee

Now the rate at which energy is generated by viscous shear
in the disk (per unit area on each disk face) is
\be
D(R) = \frac{1}{2} \nu \Sigma \left(R \frac{\partial \Omega}{\partial
R}\right)^2 = 3 \pi \nu \Sigma \left(\frac{3GM_*}{8 \pi R^3}\right) 
\label{diss}
\ee
(e.g. Frank, King \& Raine 1985).\nocite{frank1} Together with
Equation~(\ref{bloodylong}), this yields
\be
D(R)  =  \left\{ \begin{array}{ll} 
\left(\frac{3GM_*}{8 \pi R^3}\right) 
\left\{\dot{M}_{acc}(R) 
- \dot{M}_{acc}(R_*)\left(\frac{R_*}{R}\right)^{1/2} 
+ \dot{M}_{w,total}
\left(\frac{R_*}{R}\right)^{1/2}
\left[\frac{l^2 (\xi + 2)}{(\xi + 5/2)}\right]
\left[\frac{1-\left(\frac{R}{R_*}\right)^{(\xi+5/2)}}
{\left(\frac{R_{disk}}{R_*}\right)^{(\xi+2)}-1}\right]\right\}
& \xi \neq -2,-5/2 \\ \\

\left(\frac{3GM_*}{8 \pi R^3}\right) 
\left\{\dot{M}_{acc}(R) 
- \dot{M}_{acc}(R_*)\left(\frac{R_*}{R}\right)^{1/2} 
+ \dot{M}_{w,total} 
\left(\frac{R_*}{R}\right)^{1/2}
[2 l^2]
\left[\frac{1-\left(\frac{R}{R_*}\right)^{1/2}}
{\ln{\left(\frac{R_{disk}}{R_*}\right)}}\right]\right\}
& \xi = -2 \\ \\ 

\left(\frac{3GM_*}{8 \pi R^3}\right) 
\left\{\dot{M}_{acc}(R) 
- \dot{M}_{acc}(R_*)\left(\frac{R_*}{R}\right)^{1/2} 
+ \dot{M}_{w,total}
\left(\frac{R_*}{R}\right)^{1/2}
\left[\frac{l^2}{2}\right]
\left[\frac{\ln{\left(\frac{R_*}{R}\right)}}
{1-\left(\frac{R_{*}}{R_{disk}}\right)^{1/2}}\right]\right\}
& \xi = -5/2.

\end{array} 
\right.
\label{bloodylong2}
\normalsize
\ee
for a Keplerian disk. This expression correctly reduces to the usual
result for the case of no mass loss upon setting $\dot{M}_{w,total}=0$
and noting that $\dot{M}_{acc}(R)=\dot{M}_{acc}(R_{disk})={\rm const}$
in a conservative disk. The explicit form of
Equation~\ref{bloodylong2} for a conservative disk is
\be
D_{no \; wind}(R) = \frac{3GM_{*}\dot{M}_{acc}}{8\pi R^3}\left[1-\left(\frac{R_{*}}{R}\right)^{1/2}\right].
\label{diss_nowind}
\ee

It is convenient to rewrite Equation~\ref{bloodylong2} in such a way
that the difference between it and the conservative
case, Equation~\ref{diss_nowind}, becomes more explicit. After some algebraic 
manipulation, one finds that $D(R)$ can be expressed as
\be 
D(R) = D_{no \; wind}(R) -
\frac{3GM_{*}\dot{M}_{w,total}}{8\pi R^3}X(R),
\label{bloodylong3}
\ee
where
\be
X(R)= \left\{ \begin{array}{ll} 
1- \left(\frac{R_*}{R}\right)^{1/2}-
\left[\frac{\left(\frac{R}{R_*}\right)^{\xi+2}-1}
{\left(\frac{R_{disk}}{R_*}\right)^{\xi+2}-1}\right]-
\left(\frac{R_*}{R}\right)^{1/2} 
\left[\frac{l^2 (\xi + 2)}{(\xi + 5/2)}\right] 
\left[\frac{1-\left(\frac{R}{R_*}\right)^{(\xi+5/2)}}
{\left(\frac{R_{disk}}{R_*}\right)^{(\xi+2)}-1}\right] & \xi \neq -2,-5/2  \\ \\

1- \left(\frac{R_*}{R}\right)^{1/2}-
\left[\frac{\ln\left(\frac{R}{R_*}\right)}
{\ln\left(\frac{R_{disk}}{R_*}\right)}\right]-
\left(\frac{R_*}{R}\right)^{1/2} 
\left[2 l^2\right] 
\left[\frac{1-\left(\frac{R}{R_*}\right)^{1/2}}
{\ln\left(\frac{R_{disk}}{R_*}\right)}\right] & \xi = -2 \\ \\ 

1- \left(\frac{R_*}{R}\right)^{1/2}-
\left[\frac{\left(\frac{R}{R_*}\right)^{-1/2}-1}
{\left(\frac{R_{disk}}{R_*}\right)^{-1/2}-1}\right]-
\left(\frac{R_*}{R}\right)^{1/2} 
\left[\frac{l^2}{2}\right] 
\left[\frac{\ln\left(\frac{R_*}{R}\right)}
{1-\left(\frac{R_{*}}{R_{disk}}\right)^{1/2}}\right] & \xi = -5/2 
\end{array} 
\right.
\label{X_of_r}
\ee

For reference, it is noted without explicit derivation that if 
$\dot{m}(R)$ is a delta-function centered on $R_W$, then $D(R)$ can
still be written in the form of Equation~\ref{bloodylong3}, but with
$X(R)$ now given by (c.f. Equation~2.6 in Piran 1977; note that a 
power 1/2 is missing from the second term in the $R > R_W$ part of 
his equation)

\be
X(R)= \left\{
\begin{array}{ll} 
l^2\left(\frac{R_W}{R}\right)^{1/2}-
\left(\frac{R_*}{R}\right)^{1/2} & R > R_W \\ \\
1- \left(\frac{R_*}{R}\right)^{1/2} & R < R_W.
\end{array}
\right.
\label{delta_func}
\ee

\section{The radial effective temperature profile of a steady-state,
mass-losing accretion disk} 
\label{tofr}

\indent Given the rate of viscous dissipation in the steady-state disk, 
the radial effective temperature profile $T(R)$ is usually found by
assuming that the rate at which accretion energy is released by
dissipation is balanced by radiative losses, i.e. 

\be 
D(R)_{no \; wind} = \sigma T_{eff}^4(R)
\label{oldbalance}
\ee
if the disk is optically thick. However, if a wind is driven from the
surface of the disk, some of the dissipated accretion energy may not
be radiated away to infinity, but may go instead into powering the
wind. In particular, two energy sinks need to be considered: (i) the binding
energy that must be overcome to allow matter to escape at all; (ii) 
the kinetic energy carried by the flow at infinity.
\footnote{By not explicitly accounting for the thermal energy content of
the wind, we are effectively considering only models in which all of
the wind's thermal energy is eventually converted into bulk kinetic
energy (i.e. models in which $T_{wind} \rightarrow 0$ as $r
\rightarrow \infty$). Close to the disk, the thermal wind energy must 
in any case be small compared to the binding energy unless the mass 
loss is in fact thermally {\em driven}. This requires temperatures on
the order of $T_{esc} \simeq 3 \times 10^8 (M_*/M_{\sun})(R/10^9 cm)^{-1}$~K 
near the base of the flow. Such outflows have been considered by Piran
(1977) and by Czerny \& King (1989ab).}
\nocite{czerny1,czerny2}

Strictly speaking, a self-consistent dynamical model of the full
disk~+~wind system is 
needed to properly describe the process of converting the energy 
dissipated in the disk into wind energy. However, in the
spirit of the current treatment, whose point is to elucidate the 
effects of mass loss on the disk's energy balance more generally, it 
makes sense to simply treat wind energy losses, $l_w(R)$, as a local
cooling term, i.e. 
\be
D(R) = \sigma T_{eff}^4(R) + l_w(R).
\label{energy_balance}
\ee

\indent Local approximations such as this are crude, since they do not 
allow for the possibility that the dissipated accretion energy may 
be redistributed radially before powering the wind. 
It is also worth noting that, even though
Equation~\ref{energy_balance} can be made energetically correct by
construction (see below), it is quite possible for $l_w(R)$ to have a 
strong wavelength dependence. For example, in a line-driven wind,
energy losses from the
disk's radiation field to the wind may be concentrated in a few strong 
spectral lines. In this case, estimates of the effective disk temperature
derived from the observed continuum would yield values corresponding
to $l_w(R) \simeq 0$ in Equation~\ref{energy_balance}.

Ignoring these subtleties for the moment, some general statements 
can be made about the conditions under which a wind loss term is
required in the energy balance equation at all. First, the rotational 
energy that is extracted from the disk by the outflow is insufficient
to overcome 
the binding energy of the outflowing matter unless $l^2 > 3/2$. This 
can be shown, for example, by integrating $D(R)$ over $R$ 
for the case where $\dot{m}(R)$ is a delta function
(Equations~\ref{bloodylong3} and~\ref{delta_func}) and comparing the
result to the case of a conservative disk. This yields a ``luminosity
depletion'' of (c.f. Equations~2.7 and 2.8 in Piran 1977)
\be
\Delta L = \frac{GM\dot{M}_{w,total}}{2R_*} - \left(3/2 -
l^2\right)\frac{GM\dot{M}_{w,total}}{R_W}.
\label{depletion}
\ee
If the wind material had never been accreted at all, the
depletion would have been equal to the first term in
Equation~\ref{depletion}. Thus, unless $l^2 \geq 3/2$, the disk still
{\em gains} energy from the matter that is ultimately lost to the
outflow; for $l=1$, this energy is just the binding energy of the wind
material. This result should 
not come as a surprise: for $l=1$, the outflow clearly exerts no net
torque on the accretion disk, and thus no energy is extracted from the
disk to actually power the wind. The accretion disk therefore
dissipates all of gravitational binding energy it extracted from the 
wind material as it accreted from infinity to $R_W$ prior to its 
ejection. For $l<1$, even more energy can be dissipated by the disk, 
since the outflowing material carries with it less rotational kinetic 
energy than it had at the point of ejection. Specifically, for $l=0$, 
the disk gains both the rotational kinetic energy of the wind material
($GM\dot{M}_w/2R_W$) and its potential energy ($GM\dot{M}_w/R_W$).

Thus unless an external energy source is invoked to help power the
wind, $l_w$ must, in general, contain at least an effective binding
energy term for models with $l^2 < 3/2$. After all, the 
outflowing material cannot reach infinity unless at least its 
binding energy (including any rotational energy it ``injected'' into
the disk) is returned to it. Equation~\ref{depletion} also shows,
however, that models with $l^2 > 3/2$ do extract sufficient energy
from the disk to overcome the binding energy. In fact, if $l^2 >\!> 3/2$,
the energy that is extracted from the disk is sufficient to power the
wind to potentially very high terminal velocities of $v_\infty =
\sqrt{2(l^2-3/2)} GM_*/R_W \simeq \sqrt{2} l GM_*/R_W$. This
illustrates how centrifugally-driven accretion disk winds are
ultimately powered by accretion energy and shows that explicit binding
and kinetic energy terms need not be added to the right-hand side of
Equation~\ref{oldbalance} when modeling disks that drive outflows of
this type.

With these general points in mind, the wind loss term, $l_w$, can be
conveniently parameterized as follows. The rate at which energy must
be supplied (per unit area) to wind material in order for it to
overcome its binding energy and escape from the disk is 
\be 
l_b(R) = \frac{1}{2}\dot{m}(R)v^2_K(R),
\ee
where $v_K(R)$ denotes the Keplerian velocity. Here, no distinction
has been made yet between models with different lever arms, which can
have very different {\em effective} binding energies; this will be
accounted for below by means of an efficiency factor.

Similarly, kinetic energy has to be supplied to the wind at a rate of 
\be 
l_k(R) = \frac{1}{2}\dot{m}(R) v^2_\infty(R) = \frac{1}{2}\dot{m}(R)
f^2 v^2_K(R)
\ee
per unit area, where the natural scaling $v_{\infty}(R) \propto
v_{esc}(R) \propto v_K(R)$ has been assumed, with $f$ a
constant. Here, $v_{esc}(R)$ denotes the local escape velocity at 
a given streamline foot point. 

Depending on the length of the lever arm $l$ and on whether there is
an external energy source that helps to power the wind, both $l_b$
and $l_k$ may contribute fully or in part to $l_w$. Allowing for
this, $l_w$ can be written as 
\be
l_w(R) = \eta_b l_b(R) + \eta_k l_k(R) = \frac{1}{2} (\eta_b + \eta_k
f^2) \dot{m}_w(R) v_K^2(R).
\label{windlum1}
\ee
where $\eta_b$ and $\eta_k$ are efficiency factors. If there are no
external energy sources, 
\be
\eta_b = \left\{
\begin{array}{ll} 
3 - 2l^2 & l^2 < 3/2 \\ \\
0 & l^2 > 3/2
\end{array}
\right.
\ee
and 
\be
\eta_k = \left\{
\begin{array}{ll} 
1 & l^2 < 3/2 \\ \\
1 - \frac{(l^2 - 3/2)v_{esc}^2}{v_\infty^2} = 1 - \frac{2(l^2 -
3/2)}{f^2} & l^2 > 3/2.
\end{array}
\right.
\ee

\indent Equation~\ref{energy_balance} with $l_w(R)$ given by
Equation~\ref{windlum1} is the sought-after expression for the radial
effective temperature distribution across a mass-losing accretion
disk. On substituting~(\ref{bloodylong2}) or~(\ref{bloodylong3})
into~(\ref{energy_balance}), it is easy to solve for $T_{eff}(R)$
corresponding to any consistent set of parameters. More
explicitly, the result is 
\be
\sigma T_{eff}^4(R) = \frac{3GM_{*}\dot{M}_{acc}}{8\pi
R^3}\left[1-\left(\frac{R_{*}}{R}\right)^{1/2}\right] -
\frac{3GM_{*}\dot{M}_{w,total}}{8\pi R^3}X(R) -
\frac{1}{2} (\eta_b + \eta_k f^2) \dot{m}_w(R) v_K^2(R),
\label{final_expression}
\ee
where $X(R)$ is given by Equation~(\ref{X_of_r}). The physical meaning
of the terms on the right-hand side of this expression is
straightforward: the first term is the dissipation rate in the absence
of any outflow; the second term accounts for the reduction in
dissipation that results from the loss of mass and angular momentum
from the disk to the wind; the third term represents the cooling that
results if the energy dissipated by the disk must also supply some or
all of the wind's binding 
and kinetic energies. Within the second term, in the expression for
$X(R)$, terms that involve (do not involve) the lever arm $l$ arise
because the wind acts as a sink for angular momentum (mass).

\section{Summary of the model parameters}
\label{sumpar}

\indent A total of six parameters are needed to specify how the 
presence of the accretion disk wind affects the disk's radial
effective temperature distribution:

\begin{list}{}{\leftmargin 0.in}{\labelwidth 0.2in}{\labelsep 0.0in}
\protect\item[(1)] The length of the rotational lever arm
$l$. This variable permits the most meaningful division of the
available parameter space and, in principle, allows three types of
accretion disk winds to be identified: 

\begin{list}{}{\leftmargin 0.5in}
\protect\item[(i) $l < 1$:]This corresponds to the family 
of disk winds that carry away less angular momentum than
possessed by the wind material before it left the disk surface. 
Thus each disk annulus loses a larger fraction of its mass than 
of its angular momentum to the outflow. Even though such disk winds 
provide a {\em net} angular momentum sink (unless $l=0$), they 
actually increase the {\em specific}  angular momentum of the 
remaining disk material. This family of outflows is unlikely to be 
physical and will not be considered further below.
\protect\item[(ii) $l = 1$:] This corresponds to the family of
specific angular momentum-conserving disk winds. 
This prescription should be most appropriate for radiation-driven
outflows (e.g. Murray \& Chiang 1996; Proga, Stone \& Drew 1998; 
Feldmaier \& Shlosman 1999; Feldmaier, Shlosman \& Vitello 1999).
\protect\item[(iii) $l > 1$:] This corresponds to the family of accretion
disk winds that remove both specific and net angular momentum from 
the disk. This prescription is appropriate for centrifugally-driven MHD 
winds (e.g. Blandford \& Payne 1982; Cannizzo \& Pudritz 1988; 
Emmering, Blandford \& Shlosman 1992; Pelletier \& Pudritz 1992). In these, 
the length of the lever arm corresponds to the Alf\'{e}n radius, $R_{A}$,  
out to which bead-on-a-wire type rotation on magnetic field is
(effectively) enforced.\footnote{In reality, wind material does not
co-rotate all the way out to $R_A$. However, the total specific
angular momentum carried away by such an outflow (including 
magnetic torques) is exactly $\Omega R_A^2$. So, as far as the
angular momentum -- and the disk -- are concerned, the flow
effectively co-rotates out to $R_A$ (e.g. Spruit 1996).}\nocite{spruit2}
Thus $l=R_{A}/R$ in this case. Note that, in general, thermally-driven
outflows also belong to this class (Piran~1977).
\end{list}
\protect\item[(2)] The total wind mass-loss rate, $\dot{M}_{w,total}$. 
\protect\item[(3)] The radial mass loss power law index $\xi$. 
\protect\item[(4)] The velocity parameter $f=v_\infty / v_{K}$. 
\protect\item[(5)] The fraction $\eta_b$ of the wind's binding energy 
that is provided by dissipated accretion energy. Note that for $l<1$, $\eta_b$
can be greater than unity; see Section~\ref{tofr}.
\protect\item[(6)] The fraction $\eta_k$ of the wind's kinetic 
energy at infinity that is provided by dissipated accretion energy.
\end{list}

\section{Consistency of parameter combinations}

The purely parametric approach taken above to obtain the dissipation
rate and effective temperature distributions across a steady-state,
mass-losing accretion disk does not ensure that a self-consistent
solution exists for any given set of specified parameters. This is
easily seen, for example, by noting that nothing prevents us from 
specifying the mass-loss rate to be larger than the accretion rate, even
though it is clear that a steady-state solution cannot exist in this
case. Similarly, real disk winds may not give rise to power law
mass loss distributions. Once again, a full dynamical model of
the disk+wind system would be needed to avoid such pitfalls. 
The parameterized modeling approach adopted here is nevertheless 
useful because it provides a simple and general means for estimating 
how, and how much, different types of ``real'' outflows would affect
a disk's effective temperature distribution. Such estimates may be 
important as input to other types of kinematic modeling efforts
(e.g. \pcite{shlos2,me6}) and can aid in the interpretation of
$T_{eff}(R)$ distributions derived from observations
(e.g. \pcite{horne3,rutten1}). 

One consistency check that can be performed even within the present
parametric approach is to identify energetically inconsistent
parameter sets. These reveal themselves by giving rise to 
formally negative dissipation rates and effective temperatures 
in some or all disk regions, and, strictly speaking, all such
solutions should be rejected outright. However, some parameter sets
may only just fail to achieve energetic consistency, for instance as a
result of the turnover in the (standard) dissipation rate distribution
in the innermost disk regions, which the adopted power law mass loss 
distribution fails to follow. It is 
therefore useful to define a quantitative measure of the degree of
energetic inconsistency exhibited by a given solution and to consider
a somewhat less restrictive rejection criterion based on this. 

One such measure is the ratio, ${\mathcal{R}}$, of the ``negative
luminosity'' contained in any disk regions in which the dissipation
rate and/or effective temperature is formally negative to the positive
luminosity that remains 
in all other disk regions. The absolute value of this ratio is the 
fraction of the remaining luminosity that would have to be diverted to
the negative temperature regions in order to make the solution
consistent there. We may then choose to accept solutions for which
$|{\mathcal{R}}| \neq 0$, provided that $|{\mathcal{R}}| < \! < 1$. The 
plausibility argument
behind this more lenient rejection criterion is that only a minor
adjustment of the disk structure would probably be needed for such a 
solution to become consistent and that the (non-negative) dissipation
rates and effective temperatures across the disk would only have to 
change slightly to support this adjustment. For all parameter 
combinations examined in the following section $|{\mathcal{R}}| < 
0.03$, and for most combinations $|{\mathcal{R}}| = 0$ exactly.

\section{Mass-losing accretion disks in non-magnetic cataclysmic 
variables}
\label{cvs}

In this section, the formalism developed above will be used to
consider the effect of mass loss in the form of radiation-driven
(Section~\ref{rad-drive}) and centrifugally-driven winds
(Section~\ref{mag-drive}) on the effective temperature distribution of
accretion disks in cataclysmic variable stars (CVs). These two types of
outflows show the most promise for explaining the observed mass loss
in these systems.

Kinematic modeling of wind-formed, ultraviolet line profiles in CVs 
suggests mass-loss rates on the order of a few percent of the
accretion rate (Shlosman, Vitello \& Mauche 1997; Knigge \& Drew 
1997; Prinja \& Rosen 1995; Hoare \& Drew 1993). Guided by this, a
value of $\dot{M}_{w,total}/\dot{M}_{acc}(R_{disk})=0.025$ will be
adopted for 
both dynamical pictures below. Note that the current generation of
numerical and analytical models for line-driven winds is still
struggling to achieve such high efficiencies (Proga, Drew \& Stone
1998; Feldmaier \& Shlosman 1999; Feldmaier, Shlosman \& Vitello
1999). Centrifugally-driven disk winds do not share this problem, but 
do require the presence of a sufficiently strong, large-scale, ordered 
magnetic field threading the disk (Section~\ref{mag-drive}; Cannizzo
\& Pudritz 1988; Pelletier \& Pudritz 1992).\nocite{hoare2,prinja3}

For the mass loss power law index $\xi$, values of 0,-1,-2,-5/2 and -3
will be considered for all models. Note that not all of these values
are equally plausible in the context of a given dynamical
picture. This point will be considered in more detail in the relevant
sections below. Appropriate choices for the remaining outflow
parameters ($f$, $\eta_k$ and $\eta_w$) are also made and explained in
these sections.

Finally, the following system parameters are adopted throughout:
$\dot{M}_{acc}(R_{disk})=10^{-8}M_{\sun} {\rm yr}^{-1}$, $M_* = 0.5
M_{\sun}$ and $R_* = 0.0133R_{\sun}$. These are appropriate for a
non-magnetic CV in a high state, e.g. a nova-like variable like
UX~UMa.

\subsection{Radiation-driven disk winds} 
\label{rad-drive}

As already noted above, radiation-driven disk winds are expected to
belong to the $l=1$ family of models. The terminal velocities in these
types of outflows are typically on the order of the (local) escape
speeds on the disk (e.g. Proga, Stone \& Drew 1998; Feldmaier,
Shlosman \& Vitello 1999), so it is reasonable to take $f=\sqrt{2}$.

Two extreme possibilities are considered for the efficiency parameters
$\eta_b$ and $\eta_k$:

\begin{list}{}{\leftmargin 0.1in}
\protect\item[(i) {\em Minimal models}]$\left[ \eta_b=\eta_k=0 \right] $: 
For this choice of parameters, the effect of the outflow on the disk is
minimized. The minimal model is appropriate if the wind is powered
predominantly by an external energy source, such as the boundary layer
between the disk and the central star. However, somewhat
counter-intuitively, it may also be relevant if the disk does power
the wind. This is because a 
radiation-driven outflow is quite unlikely to drain energy from the 
disk's radiation field in a manner that would observationally mimic a 
cooler (blackbody) disk. Most importantly, in a line-driven wind, the
transfer of momentum and energy from the radiation field to the wind 
often takes place in a few strong lines. In this case, the overall
continuous disk spectrum is hardly affected and the minimal model is
likely to provide the most relevant benchmark for comparisons with 
observations.
\footnote{It is worth noting that, even in a hypothetical ``grey''
wind (i.e. a radiation-driven outflow with frequency-independent
opacity), energy is extracted from the radiation field by red-shifting 
the entire incident spectrum. This is not, in general, equivalent to 
(or closely approximated by) replacing the incident spectrum by one
that corresponds to a lower temperature. An excellent discussion of
momentum and energy transfer in radiation-driven stellar winds is
given by Gayley, Owocki \& Cranmer (1995).\nocite{gayley1}}
\vspace{1.5mm}
\protect\item[(ii) {\em Maximal models}]$\left[ \eta_b=\eta_k=1 \right] $:
For this choice of parameters, the effect of the outflow on the disk is
maximized. Energetically, the maximal model corresponds to the
assumption that the disk alone is responsible for powering the
outflow. However, as just noted above, the minimal model may
nevertheless be preferable to describe the observational effects of 
a purely disk-driven wind. The maximal model is included mainly to 
set an upper limit on the effect of a radiation-driven wind on the
disk's temperature distribution.
\end{list}

The radial effective temperature distributions corresponding to the 
minimal and maximal models are shown in Figures~\ref{fig:demo_temp1}
and \ref{fig:demo_temp2}. In all of these models, the presence of the
outflow results in reduced temperatures over the full range of radii,
which simply reflects the removal of mass and angular momentum: 
the mass-losing disk needs to exert less viscous torque than a 
conservative to transport material towards the accretor. As a result, 
the viscous dissipation rates and effective temperatures are also 
lower. 

\indent In most cases, the differences between $T_{eff}(R)$
with and without mass loss are small everywhere, with \\
$-\Delta T = - \left(T_{eff} - T_{eff;no\;wind}\right)
\ltappeq 10^3$~K. Such minor deviations are not observable
with current instruments and techniques. The only exceptions are the
$\eta_b=\eta_k=1$, $\xi \leq -2$ models. For these parameter
combinations, the effective temperature distribution drops
precipitously just beyond the stellar surface. The reason for this
behaviour is easy to understand. In these models, much or most of the 
mass lost to the wind is carried away from the inner disk, where the
Keplerian and escape velocities are highest. Thus the wind binding
and kinetic luminosities are largest in these cases, and, for
$\eta_b=\eta_k=1$, must be supplied entirely by the disk. As a result,
most of the accretion energy that is liberated in the innermost disk
regions is needed to power the outflow.

\indent How reasonable are these $\eta_b=\eta_k=1$, $\xi \leq -2$
parameter combinations? Intuitively, it seems inevitable that, in a
radiation-driven accretion disk wind, mass loss should occur 
preferentially from the hot, luminous, inner disk regions. Dynamical
models appear to confirm this basic expectation (Proga, Stone \& Drew
1998; Feldmaier, Shlosman \& Vitello 1999). It is particularly worth 
noting that, in a conservative disk (and, to an equally good
approximation, also in the mass-losing disks considered here),
$T_{eff} \propto R^{-3/4}$ for $R>\!>R_{*}$. This implies that 
$\xi = -3$ corresponds to the
plausible situation in which the mass loss rate per unit area is
proportional to the rate at which energy is dissipated and radiated
locally. In fact, the analytic model of Feldmaier, Shlosman \& Vitello
(1999) predicts $\dot{m}_w(R) \propto R^{-2.9}$ for $R\gtappeq 5R_{WD}$
(and $\dot{m}_w(R) \propto R^{-1}$ closer to disk centre). Thus 
$\xi \leq -2$ is a reasonable choice for a line-driven disk wind.

\indent However, as noted in Section~\ref{tofr}, the 
$\eta_b=\eta_k=1$ maximal model is likely to overestimate the
effects of radiation-driven mass loss on the disk's continuum
spectrum and on the effective temperature distribution that
would be inferred for such a disk from observations. Moreover, since 
deviations from the standard case are so highly localized in the
innermost disk regions in the $\xi=-3$ case, it is highly unlikely
that even this model would be observationally distinguishable from a
conservative disk. Even the most promising technique to infer $T(R)$ from
observations, which relies on the inversion of eclipse 
light curves of high-inclination systems (Horne 1985), cannot
currently provide the high spatial resolution that would be required 
to make such a distinction ($\Delta R <\!<1~R_{*}$). A final problem is
that an optically thick, hot boundary layer between the inner disk and
the stellar surface could effectively mask this extremely narrow break
in the disk's temperature distribution close to the WD \cite{popham1}.

Thus radiation-driven accretion disk winds are unlikely to leave an
observable imprint on the disk's effective temperature distribution.

\subsection{Centrifugally-driven disk winds} 
\label{mag-drive}

The most important parameter that must be fixed to describe a
centrifugally-driven outflow within the present formalism is the
length of the wind's effective lever arm $l$. As noted in
Section~\ref{sumpar}, in this type of outflow $lR$ corresponds to 
the Alfv\'{e}n radius, $R_A$. 

In the absence of external energy sources, a firm lower bound on 
$l$ is obtained by noting that $l^2>3/2$ is required for such a wind 
to even overcome its initial binding
energy (Section~\ref{tofr}). A rough upper bound follows from the
fact that a centrifugally-driven wind alone -- i.e. without any help
from viscous torques -- can, in principle, drive mass accretion at a
rate of about (e.g. Cannizzo \& Pudritz 1988; Pelletier \& Pudritz
1992)
\be
\frac{\dot{M}_{acc}}{\dot{M}_{wind}} \simeq
\left(\frac{R_A}{R}\right)^2 = l^2.
\label{wind-drive}
\ee
For $\dot{M}_{w,total}/\dot{M}_{acc}(R_{disk})=0.025$, this
corresponds to $l=6.3$. This is an upper limit, because observations
of disk radius changes in erupting dwarf nova systems indicate that
viscous angular momentum transport must be significant and is probably
dominant in CVs (e.g. Livio 1999). One arrives at the 
same conclusion by noting that a purely wind-driven disk, in which no
angular momentum is transported by viscosity and instead all is
extracted by an outflow, would actually be dark. This was most
recently pointed out by Spruit (1996) and also follows trivially from
the fact that there can be no viscous dissipation without viscosity: 
with $\nu=0$ in Equation~(\ref{diss}), $D(R)$ and $T(R)$ must also
vanish. This upper bound on $l$ is only approximate, mainly because
boundary effects have been ignored in the derivation of
Equation~\ref{wind-drive}. With both of these limits in mind, $l=5$ is
adopted in this section to illustrate the effect of a
centrifugally-driven wind on the disk.

Before presenting the results for models of this type, a few preliminary 
remarks are in order. First, it is of interest to consider the
strength of the large-scale, poloidal, magnetic field that is required
by centrifugally-driven outflows. Following Cannizzo \& Pudritz
(1988), one estimate can be obtained by assuming equipartition between
the magnetic and gas pressures in the disk photosphere, i.e. 
\be 
\frac{B_{eq}^2}{8\pi} = \rho c_s^2
\ee
where $\rho$ is the mass density and $c_s$ is the sound speed. Using
the standard $\alpha$-disk solutions (e.g. Frank, King \& Raine
1992\nocite{frank1}), one finds equipartition field strengths of
$B_{eq} \simeq 2000 (R/10R_{WD})^{21/16}$~G for the system parameters
adopted here. This estimate is probably close to an upper
limit: even though equipartition of gas and magnetic pressures may be
a good approximation in the disk atmosphere, it is not necessary for
all of the magnetic flux to be in the form of the large-scale field 
that is needed to launch the wind. An estimate of the minimum strength
required for {\em this} field (denoted $B_p$) is roughly given by
\be
\frac{B_p^2}{B_{eq}^2} \sim l \alpha \frac{\dot{M}_{wind}}{\dot{M}_{acc}}\frac{H}{R},
\label{bfield}
\ee
where $H$ is the disk scale-height and $\alpha$ is the usual
Shakura-Sunyaev viscosity parameter (c.f. Pringle 1993). This is
larger by a factor $l$ compared to Pringle's formula because the wind
velocity has been taken to be of order $l$-times the Keplerian
velocity.\nocite{pringle3} Equation~\ref{bfield} shows that the 
large-scale poloidal field can, 
in principle, be smaller than $B_{eq}$ by factors of 10-100.
\footnote{Note that even if the strength of the large-scale field were
close to $B_{eq}$, its presence would not imply that such a system
would be classified as a ``magnetic CV''. The latter term is reserved
for systems in which the disk is disrupted, or prevented from forming, 
by the magnetic field of the WD primary. For reference: a WD dipole 
field with a surface strength of about $5 \times 10^6$~G would be 
required to produce a field comparable to $B_{eq}$ at $10R_{WD}$, 
(taking the same system parameters as above). A system harbouring such 
a WD would indeed not be a non-magnetic CV. If a disk could form at
all, it would be truncated at a radius of about $10R_{WD}$. Even though 
this is not the type of model considered here, the interaction region 
between a stellar magnetic field and a truncated accretion disks can
itself be the launch pad for a different type of centrifugally-driven 
outflow, namely Shu~et al.'s (1994) X-celerator wind.\nocite{shu1}}

Second, it should be stressed that the origin of the large-scale
magnetic field, as 
well as the mechanism for maintaining it (with a topology suitable for
driving an outflow), have not been -- and need not be -- specified in
the present formalism. However, it is important to keep in mind that
these issues are central to the question of whether centrifugally-driven
winds can actually exist in real systems (see, for example, Tout \&
Pringle 1996; Reyes-Ruiz \& Stepinski 1996; Lubow, Papaloizou \&
Pringle 1994ab; van Ballegooijen
1989).\nocite{tout2,reyes1,lubow4,lubow3,balle1}

Third, the ratio $R_A/R = l $ might, in general, itself be a 
function of radius $R$ on the disk (e.g. Pelletier \& Pudritz 1992). 
In fact, only $\xi=-2$ is consistent with $l=$~constant in a purely
wind-driven, Keplerian disk. This is because angular momentum has to 
be removed from such a disk in just the right way to allow it to accrete 
while still remaining Keplerian. In ``mixed'' disks, such as those 
considered here, any combination of $l$ and $\xi$ is, in principle,
viable, so long as it does not imply the removal of more angular
momentum than is available. Partly because of this flexibility of
mixed disks, the additional complication involved in considering, say,
a power law form for $l$ did not seem worth pursuing here. If necessary,
it would be easy to add this, albeit at the cost of at least one
additional free parameter. Note that a mixed disk with $l=$~constant
and $\xi=-2$ should be expected to mimic a conservative disk with a
lower accretion rate.

Fourth and finally, the efficiency parameters $\eta_b$ and $\eta_k$ 
are both set to zero for the models discussed in this section. This is
because a centrifugally-driven wind already extracts enough energy
from the disk to overcome its binding energy and attain terminal
velocities well in excess of the local escape velocities 
($v_\infty \simeq l v_{esc}$; c.f. Section~\ref{tofr}).

With these remarks out of the way, the results for this class of
models are shown in Figure~\ref{fig:demo_temp3}. In all cases, the
effective disk temperatures have been affected more drastically than
in their $l=1$ counterparts. This was to be expected, since the 
outflows here carry away more angular momentum and hence remove more
accretion luminosity in the form of rotational kinetic energy. A
useful way to compare these two types of models is to note that in 
both the outflow is accelerated to about $l$ times the local escape 
velocity. Thus a centrifugally-driven wind
extracts roughly $l^2$ times more energy from the disk than even a 
maximal radiation-driven wind (unless the latter also were to attain
$v_{\infty}>\!>v_{esc}$).

The effects of a centrifugally-driven outflow on $T_{eff}(R)$ depend
on $\xi$ in a straightforward manner. Models in which mass loss does not
decline too rapidly with radius ($\xi > -2$) slightly steepen the
slope of the temperature distribution in the outer disk
region. By contrast, models with $\xi < -2$ tend to flatten
the temperature distribution further in and eventually again produce a
sharp break in the effective temperature distribution near disk
center (but note that even for the $\xi=-3$ solution $|{\mathcal{R}}|
= 0.027 <\!< 1$). For $\xi = -2$ exactly, temperatures are reduced
across the disk by an essentially constant factor.

In dynamical models of centrifugally-driven winds, $\xi$ is related to
the radial distribution of the large-scale magnetic field that threads the
disk. It is therefore hard to rule out any particular value for this
parameter {\em a priori}. The model of Blandford \& Payne (1982), for 
example, corresponds to $\xi=-2$. A potential, empirical restriction
follows from the fact that the terminal velocity of a
centrifugally-driven wind is roughly $v_{\infty} \simeq lv_{esc}$,
where $v_{esc}$ is the local escape velocity from a streamline
footpoint on the disk. If a significant fraction of the wind
escapes from the inner disk, as in the $\xi \ltappeq -2$ models, one
might expect 
to see maximum outflow velocities of about $l$-times the escape
velocity from the central object. This is somewhat higher than
observed in CVs, where the maximum velocities inferred from
wind-formed spectral lines are closer to the escape velocity itself
(e.g. Prinja \& Rosen 1995). However, it would be premature to
conclude that such models are therefore ruled out in CVs. This is 
because projection effects will tend to reduce the observed maximum
velocities. In addition, the outflow might be too dilute to
produce much line emission or absorption by the time it reaches its
terminal speed. Both effects are seen, for example, in the kinematic 
disk wind model developed by Knigge \& Drew (1997) for the nova-like
variable UX~UMa.

Figure~3 shows that only models with $\xi < -2$ are likely to be
observationally distinguishable from a conservative disk. (For
$\xi=-2$ the net change in the effective temperatures may be large
enough, but the temperature distribution closely resembles that 
of a conservative disk with a lower accretion rate.) A CV disk
that drives such a centrifugally-driven wind would have a
redder overall spectrum and would produce non-standard eclipse light 
curves if observed in a high-inclination system.
It is interesting to note that such signatures may already have been
seen in CVs. Spectroscopically, \scite{long2} showed that a
``missing'' or cooler than expected inner disk could remove the 
discrepancies between {\em Hopkins Ultraviolet Telescope} observations
of the nova-like variable IX~Vel and model accretion disk
spectra. Moreover, IX~Vel is not unique in that model disk spectra
provide a poor match to its ultraviolet spectrum -- standard disk
models tend to fail in much the same way for other high-state CVs,
both nova-likes and dwarf novae in outburst
\cite{wade1,me5,me7}. Similarly, eclipse maps derived from
observations of some high-state CVs, most notably the so-called SW~Sex
stars, suggest that the effective (really: brightness) temperature
distribution in the inner disk regions of these systems is
considerably flatter than expected \cite{rutten1}.

Unfortunately, it is far from clear that these observational
signatures are really due to the presence of an outflow, and a number 
of alternative possibilities -- e.g. magnetic disk truncation,
additional radiation sources in the system or unaccounted-for 
radiative transfer effects -- are discussed 
in the articles cited above. Nevertheless, given that CVs (and other 
disk accretors) are {\em known} to drive accretion disk winds,
attempts to explain discrepancies between standard disk models and
observations in terms of the effects of these winds certainly have
some appeal.

\section{Conclusions}

The radial distributions of dissipation rate and effective temperature
across a Keplerian, steady-state, mass-losing accretion disk have been
derived and examined. The parametric approach used to achieve 
this is sufficiently general to be applicable to many types of
dynamical disk wind models, including both radiation-driven and
centrifugally-driven outflows. 

To simplify the calculations, the local wind mass-loss rate was
assumed to follow a power law distribution in radius. In addition,
wind material moving along a streamline originating at radius $R$ in
the disk was assumed to effectively co-rotate with the disk out to a radial
distance $lR$, so that non-rotating ($l=0$), specific angular
momentum-conserving ($l=1$) and angular momentum-extracting ($l>1$)
disk winds could all be described within the same framework. To
account for cases where the wind is partially or entirely
powered by energy dissipated within disk, fixed (but arbitrary)
fractions of the wind's binding and kinetic luminosities were used as
cooling terms in deriving the disks effective temperature
distribution. In the absence of external (non-disk) energy sources, 
the energetics of the disk+wind system constrain these fractions 
to be simple functions of the length of the lever arm $l$. With these
assumptions and simplifications, the mass-losing disk's effective
temperature profile could be obtained analytically.

Effective temperature distributions for CV accretion disks that lose
mass to radiation-driven or centrifugally-driven winds have been
calculated and discussed in detail. For realistic wind mass-loss rates
of a few percent centrifugally-driven outflows -- 
particularly those in which mass loss is concentrated near disk centre
-- appear to significantly alter the disk's effective 
temperature distribution. An accretion disk that is affected by the
presence of such an outflow could produce spectra and eclipse light
curves that are noticeably different from those produced by a
standard, conservative disk. In fact, existing discrepancies between 
disk models and observations of some systems could probably be
reconciled within this framework. 
However, since other mechanisms might also still be able to
account for these discrepancies, the latter cannot yet be regarded 
as clear signatures of a disk wind. This uniqueness problem will
disappear once a self-consistent and reliable dynamical picture of
accretion disk winds in CVs and other disk accreting systems has been
found. This will allow the free parameters of the parametric disk wind
model developed here to be fixed. The formalism outlined above can
then be used to obtain a simple, but reliable estimate of the impact
of the wind on the disk's effective temperature distribution.

\section*{Acknowledgments}
The author would like to thank Luis Tao, Mike Goad, Knox Long, 
Mario Livio and Joe Patterson for a number of helpful discussions and 
suggestions related to this work. The comments of an anonymous referee 
also led to substantial improvements and are gratefully acknowledged. 
Support was provided in part by NASA through Hubble Fellowship grant HF-01109
awarded by the Space Telescope Science Institute, which is operated by the
Association of Universities for Research in Astronomy, Inc., for NASA
under contract NAS 5-26555. This work is Columbia Astrophysics Laboratory 
contribution number 687.

\bibliographystyle{mnras}
\bibliography{bibliography}

\begin{thebibliography}{Emmering, Blandford \& Shlosman<1992>}

\bibitem[Bath, Edwards \& Mantle<1983>]{bath4}
Bath, G.~T., Edwards, A.~C.  \& Mantle, V.~J., 1983.
\newblock {\it MNRAS}, {\bf 205}, 171.

\bibitem[Blandford \& Payne<1982>]{bland1}
Blandford, R.~D. \& Payne, D.~G., 1982.
\newblock {\it MNRAS}, {\bf 199}, 883.

\bibitem[Cannizzo \& Pudritz<1988>]{cann1}
Cannizzo, J.~K. \& Pudritz, R.~E., 1988.
\newblock {\it ApJ}, {\bf 327}, 840.

\bibitem[Czerny \& King<1989a>]{czerny1}
Czerny, M. \& King, A.~R., 1989a.
\newblock {\it MNRAS}, {\bf 236}, 843.

\bibitem[Czerny \& King<1989b>]{czerny2}
Czerny, M. \& King, A.~R., 1989b.
\newblock {\it MNRAS}, {\bf 241}, 839.

\bibitem[Emmering, Blandford \& Shlosman<1992>]{emmer1}
Emmering, R.~T., Blandford, R.~D.  \& Shlosman, I., 1992.
\newblock {\it ApJ}, {\bf 385}, 460.

\bibitem[Feldmaier \& Shlosman<1999>]{feld1}
Feldmaier, A. \& Shlosman, I., 1999.
\newblock {\it ApJ}, {\bf submitted}.

\bibitem[Feldmaier, Shlosman \& Vitello<1999>]{feld2}
Feldmaier, A., Shlosman, I.  \& Vitello, P. A.~J., 1999.
\newblock {\it ApJ}, {\bf submitted}.

\bibitem[Frank, King \& Raine<1992>]{frank1}
Frank, J., King, A.~R.  \& Raine, D.~J., 1992.
\newblock {\it Accretion Power in Astrophysics}, Cambridge University Press,
  Cambridge.

\bibitem[Gayley, Owocki \& Cranmer<1995>]{gayley1}
Gayley, K.~G., Owocki, S.~P.  \& Cranmer, S.~R., 1995.
\newblock {\it ApJ}, {\bf 442}, 296.

\bibitem[Hoare \& Drew<1993>]{hoare2}
Hoare, M.~G. \& Drew, J.~E., 1993.
\newblock {\it MNRAS}, {\bf 260}, 647.

\bibitem[Horne<1985>]{horne3}
Horne, K., 1985.
\newblock {\it MNRAS}, {\bf 213}, 129.

\bibitem[Knigge \& Drew<1997>]{me6}
Knigge, C. \& Drew, J.~E., 1997.
\newblock {\it ApJ}, {\bf 486}, 445.

\bibitem[Knigge {\it et~al.}<1997>]{me5}
Knigge, C., Long, K.~S., Blair, W.~P.  \& Wade, R.~A., 1997.
\newblock {\it ApJ}, {\bf 476}, 291.

\bibitem[Knigge {\it et~al.}<1998>]{me7}
Knigge, C., Long, K.~S., Wade, R.~A., Baptista, R., Horne, K., Hubeny, I.  \&
  Rutten, R. G.~M, 1998.
\newblock {\it ApJ}, {\bf 499}, 414.

\bibitem[Livio<1999>]{livio7}
Livio, M., 1999.
\newblock {\it Physics Reports}, {\bf 311}, 225.

\bibitem[Long {\it et~al.}<1994>]{long2}
Long, K.~S., Wade, R.~A., Blair, W.~P., Davidsen, A.~F.  \& Hubeny, I., 1994.
\newblock {\it ApJ}, {\bf 426}, 704.

\bibitem[Lubow, Papaloizou \& Pringle<1994a>]{lubow4}
Lubow, S.~H., Papaloizou, J. C.~B  \& Pringle, J.~E., 1994a.
\newblock {\it MNRAS}, {\bf 267}, 235.

\bibitem[Lubow, Papaloizou \& Pringle<1994b>]{lubow3}
Lubow, S.~H., Papaloizou, J. C.~B  \& Pringle, J.~E., 1994b.
\newblock {\it MNRAS}, {\bf 268}, 1010.

\bibitem[Murray \& Chiang<1996>]{murray2}
Murray, N. \& Chiang, J., 1996.
\newblock {\it Nat}, {\bf 382}, 789.

\bibitem[Pelletier \& Pudritz<1992>]{pell1}
Pelletier, G. \& Pudritz, R.~E., 1992.
\newblock {\it ApJ}, {\bf 394}, 117.

\bibitem[Piran<1977>]{piran1}
Piran, T., 1977.
\newblock {\it MNRAS}, {\bf 180}, 45.

\bibitem[Popham \& Narayan<1995>]{popham1}
Popham, R. \& Narayan, R., 1995.
\newblock {\it ApJ}, {\bf 442}, 337.

\bibitem[Pringle<1981>]{pringle1}
Pringle, J.~E., 1981.
\newblock {\it ARA\&A}, {\bf 19}, 137.

\bibitem[Pringle<1992>]{pringle3}
Pringle, J.~E., 1992.
\newblock In: {\it {Astrophysical Jets}}, p. 1, eds Burgarella, D., Livio, M.
  \& O'Dea, C.~P.

\bibitem[Prinja \& Rosen<1995>]{prinja3}
Prinja, R.~K. \& Rosen, S.~R., 1995.
\newblock {\it MNRAS}, {\bf 273}, 461.

\bibitem[Proga, Stone \& Drew<1998>]{proga1}
Proga, D., Stone, J.  \& Drew, J.~E., 1998.
\newblock {\it MNRAS}, {\bf 295}, 595.

\bibitem[Reyes-Ruiz \& Stepinski<1996>]{reyes1}
Reyes-Ruiz, M. \& Stepinski, T.~F., 1996.
\newblock {\it ApJ}, {\bf 459}, 653.

\bibitem[Rutten, vanParadijs \& Tinbergen<1992>]{rutten1}
Rutten, R. G.~M., van Paradijs, J.  \& Tinbergen, J., 1992.
\newblock {\it A\&A}, {\bf 260}, 213.

\bibitem[Shlosman, Vitello \& Mauche<1996>]{shlos2}
Shlosman, I., Vitello, P. A.~J.  \& Mauche, C.~W., 1996.
\newblock {\it ApJ}, {\bf 461}, 377.

\bibitem[Shu {\it et~al.}<1994>]{shu1}
Shu, F., Najita, J., Ostriker, E., Wilkin, F., Ruden, S.  \& Lizano, S., 1994.
\newblock {\it ApJ}, {\bf 429}, 781.

\bibitem[Spruit<1996>]{spruit2}
Spruit, H.~C., 1996.
\newblock In: {\it {Evolutionary Processes in Binary Stars, NATO ASI Series C,
  Vol. 477}}, p 249, eds Wijers, R. A. M.~J., Davies, M.~B.  \& Tout, C.~A.,
  Kluwer Dordrecht.

\bibitem[Tout \& Pringle<1996>]{tout2}
Tout, C.~A. \& Pringle, J.~E., 1996.
\newblock {\it MNRAS}, {\bf 281}, 219.

\bibitem[van Ballegooijen<1989>]{balle1}
van Ballegooijen, A.~A., 1989.
\newblock In: {\it Accretion Disks and Magnetic Fields in Astrophysics}, p. 99,
  ed. Belvedere, G., Kluwer, Dordrecht.

\bibitem[Wade<1988>]{wade1}
Wade, R.~A., 1988.
\newblock {\it ApJ}, {\bf 335}, 394.

\end{thebibliography}

\pagebreak
 
\begin{figure*}
\centering
\begin{picture}(600,500)
\put(0,0){\includegraphics{temp_m0025_l1_eta0.eps}} 
\end{picture}
\caption[Radial effective temperature profiles of mass-losing disks:
low mass-loss rate and rotation.]
{{\em Top panel:} Radial effective temperature distributions of disks
losing mass to a radiation-driven wind. The outflow is assumed to be 
specific angular momentum-conserving ($l=1$), with a mass-loss rate of 
$\dot{M}_{w,total}/\dot{M}_{acc}(R_{disk})=0.025$. None of the wind's
binding or mechanical luminosity is provided by dissipated accretion 
energy in these ``minimal'' models ($\eta_b=\eta_k=0$). Each line 
corresponds to a different distribution of the local mass-loss rate 
per unit area with radius, $\dot{m}_{wind}(R) \propto R^{-\xi}$, as
indicated. 
{\em Bottom panel:} The effective temperature differences $-\Delta T =
- \left[ T_{eff} - T_{eff;no\;wind}\right]$  between the mass-losing
disk models and a conservative disk.\label{fig:demo_temp1}}
\end{figure*}

\begin{figure*}
\centering
\begin{picture}(600,500)
\put(0,0){\includegraphics{temp_m0025_l1_eta1.eps}} 
\end{picture}
\caption[Radial effective temperature profiles of mass-losing disks:
low mass-loss rate and rotation.]
{{\em Top panel:} Radial effective temperature distributions of disks
losing mass to a radiation-driven wind. The outflow is assumed to be 
specific angular momentum-conserving ($l=1$), with a mass-loss rate of 
$\dot{M}_{w,total}/\dot{M}_{acc}(R_{disk})=0.025$. All of the wind's
binding or mechanical luminosity is provided by dissipated accretion 
energy in these ``maximal'' models ($\eta_b=\eta_k=1$). Each line 
corresponds to a different distribution of the local mass-loss rate 
per unit area with radius, $\dot{m}_{wind}(R) \propto R^{-\xi}$, as
indicated. 
{\em Bottom panel:} The effective temperature differences $-\Delta T =
- \left[ T_{eff} - T_{eff;no\;wind}\right]$  between the mass-losing
disk models and a conservative disk.\label{fig:demo_temp2}}
\end{figure*}

\begin{figure*}
\centering
\begin{picture}(600,500)
\put(0,0){\includegraphics{temp_m0025_l5_eta0.eps}} 
\end{picture}
\caption[Radial effective temperature profiles of mass-losing disks:
low mass-loss rate and rotation.]
{{\em Top panel:} Radial effective temperature distributions of disks
losing mass to a centrifugally-driven wind. The outflow extracts
angular momentum from the disk ($l=5$) and has a mass-loss rate of  
$\dot{M}_{w,total}/\dot{M}_{acc}(R_{disk})=0.025$. None of the wind's
binding or mechanical luminosity needs to be provided by dissipated
accretion energy in these models ($\eta_b=\eta_k=0$), since the
outflow already extracts sufficient energy from the disk to accelerate
the ejected material to velocities in excess of the local escape
velocity. Each line corresponds to a different distribution of the
local mass-loss rate per unit area with radius, $\dot{m}_{wind}(R)
\propto R^{-\xi}$, as indicated.
{\em Bottom panel:} The effective temperature differences $-\Delta T =
- \left[ T_{eff} - T_{eff;no\;wind}\right]$  between the mass-losing
disk models and a conservative disk.\label{fig:demo_temp3}}
\end{figure*}

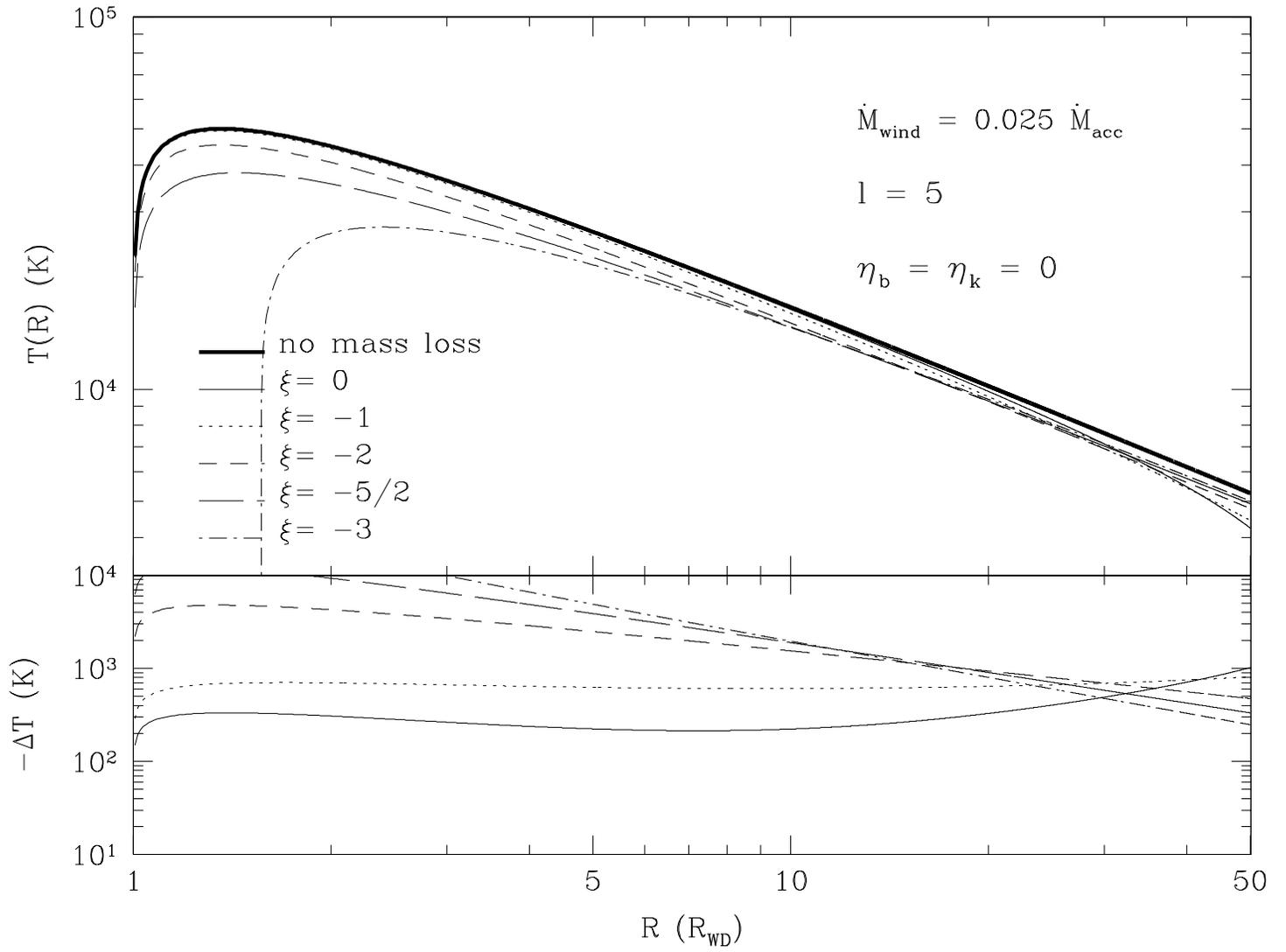
\end{document}